\documentclass[11pt]{article}

\begin{document}

\begin{center}
{\Large\bf{}The gravitational angular momentum for the super-energy Bel-Robinson tensor}
\end{center}

\begin{center}
Lau Loi So
\end{center}

\begin{abstract}
Although the super-energy Bel-Robinson tensor gives a desirable gravitational energy-momentum in a small sphere region, the angular-momentum is vanishing.  Intuitively, it should be non-zero. Our present work shows that indeed the angular momentum is non-vanishing under the continuity equation requirement. Meanwhile  this angular momentum can be converted as a ``Poynting" vector. In addition, by the analog, we also constructed the four ``Maxwell" equations in general relativity. 
\end{abstract}

\section{Introduction}
Since 1958, Bel and Robinson started study a 4th rank tensor to describe the gravitational energy-momentum~\cite{Bel19581st,Bel19582nd,Robinson19581st,Robinson19972nd}.  The dimensional difficulty can be solved by the method of a small sphere limit~\cite{Horowitz,Szabados}.  Moreover, Senovilla written a nice paper for this Bel-Robinson tensor in general relativity~\cite{Senovilla}. Inside a small sphere region, the Be-Robinson tensor gives a nice energy-momentum values, but the angular momentum is vanishing.  Somehow accepting that zero angular momentum is counterintuitive.

The reason why the angular momentum cannot have an appropriate value may be just relied on the static case: the lowest order electric part $E_{ab}$ and magnetic part $B_{ab}$ for the Riemann curvature tensor.  In fact, these two components $(E_{ab},B_{ab})$ are time dependent, for instance the tidal heating energy dissipation phenomenon~\cite{Thorne,Zhang,Poisson}. Thus one can treat the angular momentum under the time varying scenario $\dot{E}_{ab}$ and $\dot{B}_{ab}$.  Here we claim that using $(\dot{E}_{ab},\dot{B}_{ab})$ and inside a small sphere region, the angular momentum for the Bel-Robinson tensor is non-vanishing.  In addition, our result can be interpreted as a ``Poynting" vector. Meanwhile, by the analog, we also constructed the four ``Maxwell" equations in general relativity.

\section{Technical background}

Throughout this work, we use the same spacetime signature and
notation in MTW~\cite{MTW}, including the geometrical units
$G=c=1$, where $G$ and $c$ are the Newtonian constant and the speed of light. The Greek letters denote the spacetime and Latin letters refer to spatial. In vacuum, the Bel-Robinson tensor can be defined as follows
\begin{eqnarray}
B_{\alpha\beta\mu\nu}:=R_{\alpha\lambda\mu\sigma}R_{\beta}{}^{\lambda}{}_{\nu}{}^{\sigma}
+R_{\alpha\lambda\nu\sigma}R_{\beta}{}^{\lambda}{}_{\mu}{}^{\sigma}
-\frac{1}{8}g_{\alpha\beta}g_{\mu\nu}R^{2}_{\lambda\sigma\rho\tau},\label{6aJan2024}
\end{eqnarray}
where
$R^{2}_{\lambda\sigma\rho\tau}=R_{\lambda\sigma\rho\tau}R^{\lambda\sigma\rho\tau}$.
In order to extract the energy-momentum, one can
use the analog of the ``electric" $E_{ab}$ and ``magnetic"
$B_{ab}$ parts of the Weyl tensor \cite{Carmeli},
\begin{eqnarray}
E_{ab}:=C_{0a0b}, \quad B_{ab}:=\ast{C_{0a0b}},
\end{eqnarray}
where $C_{\alpha\beta\mu\nu}$ is the Weyl conformal tensor and
$\ast{C_{\alpha\beta\mu\nu}}$ is its dual,
\begin{equation}
\ast{C_{\alpha\beta\mu\nu}}=\frac{1}{2}
\epsilon_{\alpha\beta\lambda\sigma} C^{\lambda\sigma}{}_{\mu\nu}.
\end{equation}
(Here
$\epsilon_{\alpha\beta\mu\nu}=\epsilon_{[\alpha\beta\mu\nu]}$ with
$\epsilon_{0123}=\sqrt{-g}$ is the totally anti-symmetric
Levi-Civita tensor, see \cite{MTW}, in particular Eq. 8.10 and Ex.
8.3.) In vacuum using the Riemann tensor
\begin{equation}
E_{ab}=R_{0a0b},\quad B_{ab}=*R_{0a0b}.
\end{equation}
Certain commonly occurring quadratic combinations of the Riemann
tensor components in terms of the electric $E_{ab}$ and magnetic
$B_{ab}$ parts in vacuum are
\begin{equation}
R_{0a0b}R_{0}{}^{a}{}_{0}{}^{b}=E^{2}_{ab},\quad
R_{0abc}R_{0}{}^{abc}=2B^{2}_{ab},\quad
R_{abcd}R^{abcd}=4E^{2}_{ab},\label{27uDec2023}
\end{equation}
where $E^{2}_{ab}:=E_{ab}E^{ab}$ and similarly for $B^{2}_{ab}$. In particular, the Riemann squared tensor can then be written
\begin{equation}
R^{2}_{\lambda\sigma\rho\tau} =8(E^{2}_{ab}-B^{2}_{ab}).\label{27vDec2023}
\end{equation}

In empty space, the three dimensional Maxwell equations are
\begin{eqnarray}
\vec{\nabla}\cdot\vec{E}=0,\quad\vec{\nabla}\cdot\vec{B}~=~0,
\quad\vec{\nabla}\times\vec{E}=-\partial_{0}\vec{B},
\quad\vec{\nabla}\times\vec{B}=\partial_{0}\vec{E},\label{8bJan2024}
\end{eqnarray}
where the electric filed $\vec{E}=(E_{1},E_{2},E_{3})$ and likewise the magnetic field $\vec{B}=(B_{1},B_{2},B_{3})$. Using $(\vec{E},\vec{B})$ as an analog in electromagnetism, we have the ``Maxwell" equations in general relativity
\begin{eqnarray}
\vec{\nabla}\cdot\vec{E_{i}}=0,\quad\vec{\nabla}\cdot\vec{B_{i}}=0,\quad\vec{\nabla}\times\vec{E_{i}}=\partial_{0}\vec{B_{i}},\quad\vec{\nabla}\times\vec{B}_{i}=-\partial_{0}\vec{E}_{i},\label{8dJan2024}
\end{eqnarray}
where $\vec{E}_{i}=(E_{i1},E_{i2},E_{i3})$ and similarly for $\vec{B}_{i}$.  
We claim that the above four analog ``Maxwell" equations can be derived by the following Bianchi identities respectively
\begin{eqnarray}
0&=&\partial_{a}R^{a}{}_{0b0}+\partial_{b}R^{a}{}_{00a}+\partial_{0}R^{a}{}_{0ab},\\
0&=&\partial_{a}R^{a}{}_{0bc}+\partial_{b}R^{a}{}_{0ca}+\partial_{c}R^{a}{}_{0ab},\\
0&=&\partial_{c}R_{0a0b}+\partial_{0}R_{0abc}+\partial_{b}R_{0ac0},\\
0&=&\partial_{d}R_{0abc}+\partial_{0}R_{adbc}+\partial_{a}R_{d0bc},
\end{eqnarray}
where the substitution may require the classification of the Weyl conformal tensor, i.e., classifying the gravitational field~\cite{Carmeli}.  Moreover, we use (\ref{8dJan2024}) to derive the continuity equation in general relativity: 
\begin{eqnarray}
0&=&\vec{E}^{i}\cdot(\partial_{0}\vec{E}_{i}
+\vec{\nabla}\times\vec{B}_{i})\nonumber\\
&=&\vec{E}^{i}\cdot(\partial_{0}\vec{E}_{i})
-\partial^{c}(\epsilon_{cab}E^{ad}B^{b}{}_{d})
+\vec{B}^{i}\cdot(\vec{\nabla}\times\vec{E}_{i})
\nonumber\\
&=&\frac{1}{2}\partial_{0}(|\vec{E}_{i}|^{2}+|\vec{B}_{i}|^{2})
-\vec{\nabla}\cdot\vec{S}_{G},
\end{eqnarray}
where the analog ``Poynting" vector in general relativity is 
\begin{eqnarray}
\vec{S}_{G}&=&\epsilon_{cab}E^{ad}B^{b}{}_{d}\nonumber\\
&=&\left[(\vec{E}_{2}\cdot\vec{B}_{3}-\vec{E}_{3}\cdot\vec{B}_{2}),\,
(\vec{E}_{3}\cdot\vec{B}_{1}-\vec{E}_{1}\cdot\vec{B}_{3}),\,
(\vec{E}_{1}\cdot\vec{B}_{2}-\vec{E}_{2}\cdot\vec{B}_{1})\right].
\end{eqnarray}
Physically, this ``Poynting" vector $\vec{S}_{G}$ measure the energy flow, but we also claim that that vector also represent a kind of gravitational radiation.  Here we rephrase the argument from Goswami and Ellis~\cite{Ellis}: Suppose $\vec{S}_{G}$ is not vanishing, then both $(E_{ab},B_{ab})$ cannot be zero simultaneously.  In contrast, presumably that either $E_{ab}$ or $B_{ab}$ vanishes, then the ``Poynting" vector $\vec{S}_{G}$ has to be vanished.  Furthermore using (\ref{8dJan2024}), we recovered the wave equations for $(\vec{E}_{i},\vec{B}_{i})$ such that the speed of gravitational wave in vacuum is indeed the speed of light:
\begin{eqnarray}
&&0=\vec{\nabla}\times(\vec{\nabla}\times\vec{E}_{i}-\partial_{0}\vec{B}_{i})
=(\partial^{2}_{0}-\vec{\nabla}^{2})\vec{E}_{i},\\
&&0=\vec{\nabla}\times(\vec{\nabla}\times\vec{B}_{i}+\partial_{0}\vec{E}_{i})=(\partial^{2}_{0}-\vec{\nabla}^{2})\vec{B}_{i}.
\end{eqnarray}

Here come to the technical detail.  In a weak field the metric tensor can be decomposed as
$g_{\alpha\beta}=\eta_{\alpha\beta}+h_{\alpha\beta}$, and its
inverse is $g^{\alpha\beta}=\eta^{\alpha\beta}-h^{\alpha\beta}$. 
We only consider the lowest order (see (8) at~\cite{Zhang}),
the metric components can be written as
\begin{eqnarray}
h^{00}&=&\frac{3}{r^{5}}I_{ab}x^{a}x^{b}-E_{ab}x^{a}x^{b},\label{27xDec2023}\\
h^{0j}&=&\frac{4}{r^{5}}\,\epsilon^{j}{}_{pq}J^{p}{}_{l}\,x^{q}x^{l}
+\frac{2}{3}\epsilon^{j}{}_{pq}B^{p}{}_{l}x^{q}x^{l}
+\frac{2}{r^{3}}\dot{I}^{j}{}_{a}\,x^{a}
+\frac{10}{21}\dot{E}_{ab}x^{a}x^{b}x^{j}
-\frac{4}{21}\dot{E}^{j}{}_{a}x^{a}r^{2},\label{27yDec2023}\\
h^{ij}&=&\eta^{ij}h^{00}+\bar{h}^{ij},\label{27zDec2023}
\end{eqnarray}
where
\begin{eqnarray}
\bar{h}^{ij}=\frac{8}{3r^{3}}\epsilon_{pq}{}^{(i}\dot{J}^{j)p}x^{q}
+\frac{5}{21}x^{(i}\epsilon^{j)}{}_{pq}\dot{B}^{q}{}_{l}x^{p}x^{l}
-\frac{1}{21}r^{2}\epsilon_{pq}{}^{(i}\dot{B}^{j)\,q}x^{p}.
\end{eqnarray}
Zhang used $\bar{h}^{\alpha\beta}$ for the manipulation while we
prefer using $h^{\alpha\beta}$, the transformation is as follows
\begin{eqnarray}
\bar{h}^{\alpha\beta}
=h^{\alpha\beta}-\frac{1}{2}\eta^{\alpha\beta}h.
\end{eqnarray}
The corresponding first order harmonic gauge is
$\partial_{\beta}\bar{h}^{\alpha\beta}=0$.  Our calculation limit find that both the mass quadrupole moment $I_{ij}$ and current quadrupole moment $J_{ij}$ are not required. But we recall the relation for $(I_{ij},\dot{E}_{ij})$ and $(J_{ij},\dot{B}_{ij})$ as determined by
Poisson~\cite{Poisson}:
\begin{eqnarray}
I^{ij}=\frac{32}{45}M^{6}\dot{E}^{ij},\quad{}
J^{ij}=\frac{8}{15}M^{6}\dot{B}^{ij},
\end{eqnarray}
where $M$ is the mass of the black hole. The value of the angular momentum flux is something like
$\epsilon_{kab}(I^{ac}E^{b}{}_{c}+\frac{4}{3}J^{ac}B^{b}{}_{c})\approx\epsilon_{kab}(\dot{E}^{ac}E^{b}{}_{c}+\dot{B}^{ac}B^{b}{}_{c})M^{6}$. This quantity does not involve the energy dissipation since the energy interaction process is reversible. We emphasis  that the energy transfer is locally.

\section{Gravitational angular momentum}
In general, the angular momentum is
\begin{eqnarray}
J^{\mu\nu}:=\int(x^{\mu}P^{\nu}-x^{\nu}P^{\mu})d^{3}x,
\end{eqnarray}
where $P^{\mu}$ is the 4-momentum. In general relativity and restricted to the Bel-Robinson tensor, the gravitational angular momentum flux can be defined as follows~\cite{MTW}
\begin{eqnarray}
J^{ij}:=
\int(x^{i}B^{0j}{}_{mn}-x^{j}B^{0i}{}_{mn})\,x^{m}x^{n}\,d^{3}x,
\end{eqnarray}
where the momentum for the Bel-Robinson tensor according to (\ref{6aJan2024}) is
 \begin{eqnarray}
B_{0imn}\,x^{m}x^{n}=2(R_{0amb}R_{i}{}^{a}{}_{n}{}^{b}
-R_{0a0m}R_{i}{}^{a}{}_{0n})\,x^{m}x^{n}.\label{6bJan2024}
\end{eqnarray} 
Integrate and using (\ref{8dJan2024}), the gravitational angular momentum flux becomes
\begin{eqnarray}
J_{k}&=&\frac{1}{16\pi}\int_{V}\epsilon_{kab}\,x^{a}B^{0b}{}_{mn}\,x^{m}x^{n}\,d^{3}x\nonumber\\
&=&\frac{1}{630}\epsilon_{kab}(E^{ac}\dot{E}^{b}{}_{c}+B^{ac}\dot{B}^{b}{}_{c})\,r^{7}\nonumber\\
&=&\frac{1}{630}(\vec{\nabla}\cdot\vec{S}_{G})\,r^{7},
\end{eqnarray}
which is non-vanishing. Meanwhile the gravitational angular momentum can be represented as a ``Poynting" vector.

For the completeness, we compute the following component for the Bel-Robinson tensor
\begin{eqnarray}
B_{ijmn}\,x^{m}x^{n}&=&2(R_{0i0m}R_{0j0n}
-R_{0ima}R_{0jn}{}^{a}-R_{0mia}R_{0nj}{}^{a}
+R_{iamb}R_{i}{}^{a}{}_{n}{}^{b})\,x^{m}x^{n}\nonumber\\
&&-\frac{1}{8}r^{2}\eta_{ij}R^{2}_{\lambda\sigma\rho\tau}.
\end{eqnarray}
Taking a volume integral
\begin{eqnarray}
\frac{1}{16\pi}\int_{V}B_{ijmn}\,x^{m}x^{n}\,d^{3}x
&=&\frac{1}{60}\left[
(E^{2}_{ab}+B^{2}_{ab})\eta_{ij}-2(E^{c}{}_{i}E_{cj}+B^{c}{}_{i}B_{cj})\right]r^{5}\nonumber\\
&&+\frac{1}{1890}\left[
11(\dot{E}^{2}_{ab}+\dot{B}^{2}_{ab})\eta_{ij}-21(\dot{E}^{c}{}_{i}\dot{E}_{cj}+\dot{B}^{c}{}_{i}\dot{B}_{cj})\right]r^{7}.\label{7aJan2024}
\end{eqnarray}
As a double check, allowing $i=j$ for (\ref{7aJan2024}) and it recovered the energy density. In detail
\begin{eqnarray}
\eta^{ij}B_{ijmn}=B^{\alpha}{}_{\alpha{}mn}-B^{0}{}_{0mn}=B_{00mn},\label{7cJan2024}
\end{eqnarray}
where we have used the properties of the Bel-Robinson tensor: completely symmetric and tracelessness. Referring to (\ref{7aJan2024}) and (\ref{7cJan2024}), we recovered the gravitational energy~\cite{SoarXiv2023}
\begin{eqnarray}
\frac{1}{16\pi}\int_{V}\eta^{ij}B_{ijmn}\,x^{m}x^{n}\,d^{3}x
=\frac{1}{60}(E^{2}_{ab}+B^{2}_{ab})\,r^{5}
+\frac{2}{315}(\dot{E}^{2}_{ab}+\dot{B}^{2}_{ab})\,r^{7}.
\end{eqnarray}

\section{Conclusion}
In general relativity and in the limit of a small sphere, the super-energy Bel-Robinson tensor is the best candidate to describe the gravitational energy-momentum. It is known that tidal heating is a kind of energy dissipation phenomenon which indicates we should not only consider the static components of the electric and magnetic parts $(E_{ab},B_{ab})$, but also the time varying components $(\dot{E}_{ab},\dot{B}_{ab})$ for the Riemann curvature tensor. If using the static $(E_{ab},B_{ab})$, the angular momentum is exactly zero because of the symmetric property in a small sphere limit. Then we change the angle point of view, using the time varying $(\dot{E}_{ab},\dot{B}_{ab})$, our present work indeed shown that the angular momentum is non-vanishing because of the continuity equation. Moreover,  this angular momentum can be translated as a ``Poynting" vector.

In addition, by the analog, we have constructed the four ``Maxwell" equations in general relativity.  Meanwhile, for the completeness, we also computed the component $B_{ijmn}\,x^{m}x^{n}$ for the Bel-Robinson tensor.

\end{document}